# Mobile-to-Mobile Video Recommendation

Padmanabha Venkatagiri Seshadri, Mun Choon Chan, Wei Tsang Ooi

National University of Singapore, Computing 1, 13 Computing Drive 117417, Singapore,
{padmanab,chanmc,ooiwt}@comp.nus.edu.sg

**Abstract.** Mobile device users can now easily capture and socially share video clips in a timely manner by uploading them wirelessly to a server. When attending crowded events, such as an exhibition or the Olympic Games, however, timely sharing of videos becomes difficult due to choking bandwidth in the network infrastructure, preventing like-minded attendees from easily sharing videos with each other through a server. One solution to alleviate this problem is to use direct device-to-device communication to share videos among nearby attendees. Contact capacity between two devices, however, is limited, and thus a recommendation algorithm, such as collaborative filtering, is needed to select and transmit only videos of potential interest to an attendee. In this paper, we address the question: which video clip should be transmitted to which user. We proposed an video transmission scheduling algorithm, called *CoFiGel*, that runs in a distributed manner and aims to improve both the prediction coverage and precision of the collaborative filtering algorithm. At each device, CoFiGel transmits the video that would increase the estimated number of positive user-video ratings the most if this video is transferred to the destination device. We evaluated *CoFiGel* using real-world traces and show that substantial improvement can be achieved compared to baseline schemes that do not consider rating or contact history.

## 1 Introduction

Mobile devices are increasingly capable in their abilities to sense, capture, and store rich multimedia data. Modern mobile devices often come with built-in high quality cameras, multi-core processors, and gigabytes of storage, making them ideal portable device to capture video content. Furthermore, the devices are often equipped with multiple wireless network interfaces, making it easy for users to upload the videos captured in a timely manner to social web sites, sharing their experience with friends and the public.

In this work, we are interested in mobile video sharing among attendees of an event. As an example, consider a scenario where many mobile users are looking at exhibits and product demonstrations in an exhibition. In a large exhibition, a single user would be able to see only a small portion of the exhibits and demonstrations available. The possibility of sharing videos of exhibition and/or demonstrations among users of similar interest would greatly enhance the user experience as the users could choose to visit the exhibition booth after watching the associated video. Another application scenario is large-scale sport events such as the Olympic Games. The competitions are typically held parallelly in different venues around a city, and a spectator cannot attend all of the competitions. Sharing of video clips among supporters from the same country attending different competitions, as they move around the city, would improve the game experience. Such video sharing scenarios are applicable to many other contexts, including malls, museums, and parades. This information sharing paradigm emphasizes both spatial locality and timeliness and is very different from archived video sharing services such as those provided by YouTube.

A straight forward approach to enable such video sharing is to have users upload the videos captured to a central server through 3G/HSPA networks. Users can then search for or browse through the uploaded videos through the server. While this approach can provide good performance in terms of delivering the right videos to the right users, it has obvious drawbacks. First, when user density is high, there is likely to be insufficient aggregate upload bandwidth for the combination of large amount of data and large number of users. Next, the use of 3G/HSPA network for upload is relatively inefficient, since the network is optimized for download. Finally, videos stored in the central server have to be downloaded to individual mobile phone for viewing and rating, further straining the 3G/HSPA network.

The approach adopted in this work is to circumvent the cellular network infrastructure and transfer videos directly, from one mobile device to another mobile device, via short range connection such as WiFi or Bluetooth. A user who has captured a video and wants to share the video just need to indicate so in the mobile application. The video is pushed to nearby devices when connections to these devices becomes available. The mobile application has a video inbox, which lists the videos received by other devices. A user can choose to watch and rate any video in its inbox. A device also serves as a relay, forwarding received videos to other devices when it comes into contact with other devices, propagating videos through a network of mobile devices.

Besides alleviating the network infrastructure bottleneck, the use of direct mobile-to-mobile communication also potentially consume less power [15]. In addition, the much shorter RTT for direct mobile-to-mobile transfer allows significantly higher throughput compared to transferring large amount of data over the Internet through the 3G/HSPA network, where the median ping latency has been observed to be almost 200ms [5].

The use of short range communication among mobile devices results in intermittent connectivity. These devices, in essence, form a delay-tolerant network (DTN). As mobile devices have limited contact time, pushing the right video to a neighboring device is especially important. Ideally, we want videos that a user is interested in to end up in its inbox within a given time period.

To determine which videos is likely to be of interests to a user, our system uses a collaborative filtering (CF) based recommender system to predict the user preference. The use of this algorithm, however, requires collection of sufficient number of user-item ratings to work. In other words, pushing a video to a user now has two purposes: for the user to watch and for the user to rate. The decision to select which video to transfer should thus consider the needs of the CF algorithm as well.

To address this challenge, we propose *CoFiGel*, a video transfer scheduler in the DTN context that integrates CF-based recommender system. CoFiGel effectively utilizes the limited contact capacity among mobile devices to filter and disseminate user-generated videos published by mobile users to other mobile users. CoFiGel is designed to achieve two objectives. First, it increases the prediction coverage, which is the ability of the algorithm to predict ratings for items. Second, it routes videos in such a way that increases the item precision, i.e, the percentage of items recommended to users that are rated positively.

We evaluate CoFiGel through trace-based simulation using DTN mobility traces (RollerNet trace [11] and San Francisco taxi trace [4]) and an user rating data set from MovieLens [3]. The RollerNet trace is an example of human mobility and the San Francisco taxi trace is an example of vehicular mobility. Our evaluation shows that CoFiGel can provide 80% more prediction coverage in comparison to the baseline algorithms, detecting at least 74% of positive ratings in the process, and delivers at least 59% more positive (liked by user) items in comparison to the baseline algorithms that do not take into account either ratings or contact history.

The rest of the paper is organized as follow. Section 2 discusses related work. Section 3 describes our mobile-to-mobile video transfer application and motivates the need for CoFiGel. CoFiGel is presented in Section 4 and is evaluated in Section 5. Finally, Section 6 concludes.

## 2 Related Work

### 2.1 Collaborative Filtering (CF)

The most prominent and popular recommendation technique that has seen extensive research and wide deployment is collaborative filtering (CF) [8]. CF techniques can be broadly categorized into memory-based or model-based. Memory-based CF (MCF) utilizes rating history of users to identify neighbourhood patterns among users or items. This pattern facilitates the prediction of ratings for hitherto unrated user-item pairs. Model-based CF uses the user ratings in conjunction with standard statistical models such as Bayesian belief nets and latent semantic model to identify patterns in the ratings of user-item pairs. The resultant model is then used to make predictions for future ratings.

There exists much research on using CF on peer-to-peer (P2P) systems. PocketLens [28] is a recommender system for portable devices that uses item-item collaborative filtering for making recommendations. It proposes a rating exchange protocol for both distributed P2P architecture and centralized server architecture, where nodes rely on a central server for storing rating information. A probabilistic model-based CF is proposed by Wang et al. [13] for a P2P network. Other related work focuses on the security and privacy aspect, including providing user incentive [10], trust of rating protocol [22] and privacy [12].

## 2.2 DTN Content Dissemination

There are many unicast DTN routing schemes designed to improve point-to-point delivery probability and/or minimize delay ([30, 29, 26, 16]). These protocols, however, do not address the issue of information dissemination.

A common problem studied in DTN content dissemination is how to maximize the freshness of dynamic content ([14, 17, 21, 18]). These contents are usually downloaded from the Internet to a subset of mobile nodes, which then distribute these contents among themselves with the objective of maximizing content freshness. Caching schemes where nodes refresh/reshuffle their cached content based on a voting process can also be exploited, as done by Ioannidis et al. [17]. In [19], predefined preferences are used to route items to users. However, preferences are static and not predicted.

Another approach for content discovery and dissemination in DTN uses tags ([23, 25]). In this approach, users are expected to declare their interests in the form of a set of keywords. These keywords are disseminated into the network. When a node encounters a user with matching interest keywords, it forwards the information with matching tags to interested users.

Unlike previous work, *CoFiGel* provides a framework to integrate MCF and DTN routing, focusing on utilizing limited contact capacities in DTN to improve rating coverage and item recall. *CoFiGel* does not assume any specific MCF algorithm. Instead, it defines an abstract model of how MCF works and how the MCF should interact with the DTN routing protocol. We are not aware of any MCF that specifically takes into account the intermittent contact capacities of mobile nodes, nor any DTN mechanism that takes into account usefulness of item transferred to improve coverage and item recall of the MCF algorithm.

# 3 Mobile-to-Mobile Video Sharing

## 3.1 Motivation

We now motivate our work by demonstrating the efficacy and advantages of mobile-to-mobile video transfer.

Mobile data usage has outgrown available bandwidth in 3G/HSPA network, resulting in severe congestion in the cellular network in some cases [1]. A popular approach to reduce such congestion is to offload data traffic to the WiFi network whenever possible [2]. Communication over WiFi also consumes less power than 3G/HSPA network (four to six times less power for file transfer [15]).

We further measure the performance of file transfer between a mobile device and a central server using 3G/HSPA network and between two mobile devices directly using WiFi. To quantify the performance of mobile-to-server transfer, we upload and then download a 14.3 MB video clip to YouTube using a HSPA network. The HSPA service provides maximum download and upload rate of 7.2 Mbps and 1.9 Mbps respectively. The average download and upload throughputs measured (average of 5 trials) are 1125.2 kbps and 57 kbps respectively. To quantify the performance of mobile-to-mobile transfer, we use two Samsung Nexus S phones that support IEEE 802.11n (link rate is 72.2 Mbps) to exchange the same video file directly over a TCP connection. The measured throughput is 22.6Mbps (average of 5 trials). The large difference in measured throughput can be attributed to the differences in link rate and RTT observed (70ms for mobile-to-server and 5.5ms for mobile-to-mobile).

Having demonstrated that transfering videos between two mobile devices directly has better performance than transfering videos over infrastructural network, we now outline our approach for mobile-to-mobile video sharing.

### 3.2 Mobile-to-Mobile Video Sharing

A user of mobile-to-mobile video sharing application can choose to share any video content stored locally on her mobile device. These videos to be shared can be captured using the device's camera, or are saved onto the mobile device previously (e.g., loaded from a computer or downloaded from the Internet previously). Once a video is marked for sharing, it is placed in a *video outbox*, becomes eligible to be transferred to other devices. When the scheduling algorithm decides to transfer this video, it is pushed to a device within communication range through WiFi or Bluetooth. A video remains in the outbox even after it is transferred, so that it can be pushed to a different device later.

A video that is successfully received at the destination device enters into a *video inbox*. Over time, multiple videos can be pushed into a device's video inbox. A user, at any time, can check her video inbox and select videos of interest to watch. After watching, the video is removed from the inbox and is stored into a *video archive* until the user decides to delete it. Video inbox has a limited buffer size, while the archive is assumed to have sufficient storage for large number of videos (not unreasonable given the current storage capacity in a typical mobile device). If the inbox is full, incoming video may replace existing video in the inbox. The replaced video is removed completely from the device.

Any video that enters into a device's video inbox or archive is also eligible to transfer to another device. In this way, once a video is captured and shared, it can be pushed to multiple devices through relaying in a typical delay-tolerant networking fashion.

A challenge in such mobile-to-mobile video sharing application is limited contact time between mobile devices. It is very likely that many more videos are captured and shared than what could end up in a device's inbox. It is thus important to transfer only videos that are likely to be of interest to users.

To faciliate predictions of which videos are of interest to the users, the application uses a recommender system that runs memory-based collaborative filtering (MCF). A user can rate any video on its mobile devices, as either like or dislike. If a user watches a video without rating it, it is assumed to be implicitly rated dislike.

Each device maintains a user-video rating matrix, which is updated either when a video is rated on the device, or when the device receives a rating matrix from another device. The rating matrix is one of the two meta-data (the other is contact history among devices) being exchanged between two devices when devices make contact with each other. Meta-data transfer always have higher priority than video data transfer. Upon the update of one or more entries in the matrix, the other entries may be updated to predict the interest-level of each user-video pair based on the MCF algorithm. Based this user-video rating matrix, items in the video inbox that are predicted to be liked by the current user are recommended to the user.

### 3.3 Memory-Based Collaborative Filtering

We now explain in more detail how MCF works. MCF is a class of recommender algorithms that is model independent and is able to capture the abstract user preference on a set of items. Typical MCF techniques have the following structure. A training data set is used to build a rating matrix consisting of ratings given for items by users. The rating matrix is used to identify the similarities between users/items and also to predict the ratings of hitherto unrated items by a given user. From this sorted list of predicted ratings, a subset of highly rated items are shown to the user. Feedback from the user for the item shown is then used to update the rating matrix. The assumption is that *users tend to behave in the same way as they behaved in the past*.

For concreteness, we will use the Cosine-based similarity metric ([27, 20, 28]) in the rest of this paper to illustrate how CoFiGel works. Cosine-based similarity is a popular item-based MCF and has been used in large scale real-world applications such as the recommendation system used by *Amazon.com*. Note that CoFiGel can also work with other MCF algorithms, such as Slope One [24]. Since MCF works for recommendation of any kind of items, we will use the term *items* in the rest of the discussions to refer to videos in our application.

In general, ratings can be represented as integer values. For simplicity, we assume that ratings are binary and are expressed as either 1 (positive/like) or 0 (negative/dislike). In computing Cosine-based similarity, unrated items are assigned ratings of 0. After a user has rated an item, the item will not be recommended to the user again.

Let $U$ and $I$ be the set of all users and items respectively and $I_u^+$ and $I_u^?$ be the set of items that are rated positive and unrated by a user $u \in U$ respectively. Let the actual rating of an item $i \in I$ for user $u$ be $r_{u,i}$. Cosine-based similarity metric computes $R_{u,i}$, the *rank* of an unrated user-item pair $(u, i)$, in the following way. First, the similarity between two items $i$ and $j$ is computed using

$$Sim(i, j) = \frac{\sum_{u \in U} r_{u,i} \cdot r_{u,j}}{\sqrt{\sum_{u \in U} r_{u,i}^2} \cdot \sqrt{\sum_{u \in U} r_{u,j}^2}} \qquad (1)$$

For each unrated item $i \in I_u^?$, $R_{u,i}$ is computed as:

$$R_{u,i} = \sum_{j \in I_u^+} Sim(i, j) \qquad (2)$$

Obviously, the rank of item $i$ for user $u$ can be computed only if there is at least one user who has rated both $i$ and another item that user $u$ has rated positively. If the rank cannot be computed, then we say that the particular user-item pair is *unpredictable*. Table 1 shows a rating matrix with items that are rated (positively and negatively), predicted and unpredictable.

Typically, the top-$k$ items $i \in I_u^?$ with highest rank are recommended to user $u$. We say that the prediction of $i$ is *positive* for $u$ if $i$ is among the top-$k$ items in $I_u^?$, and *negative* otherwise. A prediction of $i$ is said to be *correct*, if the predicted rating is consistent with the user rating eventually. Note that the notion of whether a prediction is positive or not changes over time (and thus whether it is correct or not changes over time as well).

**Table 1.** Rating matrix for Cosine-based similarity metric, $\diamond$ denotes ratings that could be predicted and $\star$ denotes unknown ratings

| Users | $i_1$ | $i_2$ | $i_3$ | $i_4$ | $i_5$ | $i_6$ |
|---|---|---|---|---|---|---|
| $u_1$ | 1 | $\diamond$ | $\star$ | $\star$ | $\diamond$ | $\diamond$ |
| $u_2$ | 1 | $\diamond$ | $\star$ | $\star$ | $\diamond$ | $\diamond$ |
| $u_3$ | 1 | 1 | $\star$ | $\diamond$ | $\diamond$ | 1 |
| $u_4$ | $\diamond$ | 1 | $\diamond$ | 1 | 1 | 1 |
| $u_5$ | $\star$ | $\diamond$ | 1 | 1 | $\diamond$ | $\diamond$ |
| $u_6$ | 1 | $\diamond$ | $\star$ | $\diamond$ | 1 | 1 |
| $u_7$ | 1 | 0 | $\star$ | $\diamond$ | 0 | 1 |

**Table 2.** Predicted rating and coverage for $(u_4, i_1)$ and $(u_4, i_3)$ user-item pairs

| User-Item | Predicted Rating | Gain in rated and predictable items |
|---|---|---|
| $(u_4, i_1)$ | 1.30 | 2 |
| $(u_4, i_3)$ | 0.71 | 4 |

The performance of MCF algorithm is measured by several standard metrics [8]. For instance, *precision* and *recall* are used to measure the classification performance of a MCF algorithm. Precision is a measure of recommended items that are relevant to the users, and recall is a measure of the number of relevant items that are recommended to the users. Another common performance measure used is *prediction coverage*, (or coverage for short), defined as the percentage of *the number of predictable user-item pair*.

### 3.4 MCF for Mobile-to-Mobile Recommendation

When two mobile devices meet, they need to select which items to be transmitted over the intermittent contacts based on the meta-data information available. As mentioned, since contact capacity is precious, items that are likely to be liked by other users should be transfer and propagated with higher priorities. Running MCF in the context of mobile-to-mobile video sharing, however, leads to another issue: since each user is likely get a chance to rate only a small subset of all videos available, selecting which items for users to rate is also important, to increase the coverage. We illustrate this intricacy in the rest of this section with an example.

Consider the rating matrix shown in Table 1. Item $i_1$ has three common user ratings with items $i_2$, $i_5$ and $i_6$. $i_3$ has only one common user rating with $i_4$. Using Equations 1 and 2, we can compute $R_{u_4, i_1}$ as,

$$R_{u_4,i_1} = Sim(1,2) + Sim(1,4) + Sim(1,5) + Sim(1,6)$$
$$= \frac{1}{\sqrt{5}\sqrt{2}} + 0 + \frac{1}{\sqrt{5}\sqrt{2}} + \frac{3}{\sqrt{5}\sqrt{4}} \approx 1.30$$

Similarly,

$$R_{u_4,i_3} = Sim(3,2) + Sim(3,4) + Sim(3,5) + Sim(3,6)$$
$$= 0 + \frac{1}{\sqrt{1}\sqrt{2}} + 0 + 0 \approx 0.71$$

The results are listed in Table 2. $i_1$ has a higher rating than $i_3$ with respect to $u_4$.

However, the consideration, in terms of coverage, is different. It can be observed from Table 1 that all users except $u_4$ and $u_5$ have already rated $i_1$. Knowing the value of $r_{u4,i_1}$, allows only at most one more rating, $R_{u_5,i_1}$, to be computed. The gain in rated and predictable items is 2. On the other hand, $i_3$ has been rated only by $u_5$. Knowing the value of $r_{u_4,i_3}$, allows the rating of 3 users ($u_3$, $u_6$ and $u_7$) for item $i_3$ to be computed. The gain in rated and predictable items is 4. Therefore, the rating of $i_3$ by $u_4$ has a higher gain in rated and predictable items than rating $i_1$.

This example illustrates the trade-off between improving user satisfaction and improving coverage when not all data transfer can be completed within a contact. If user satisfaction is more important, then $i_1$ will be chosen for transfer. If coverage has higher priority, then $i_3$ should be chosen.

Note that when there is a centralized server with continuous connectivity to users and has access to all rating information and data items, the impact of this trade-off is not significant. However, such a trade-off plays an important role in a resource constraint environment where the contacts between mobile devices are intermittent, contact capacities are limited and only subsets of data items can be stored in the local buffers.

*The execution of MCF on mobile devices with intermittent contacts presents a new challenge that is not present in traditional applicaiton of MCF in a centralized or peer-to-peer environment where connectivities are not intermittent.*

## 4 CoFiGel

### 4.1 System Model

The MCF algorithm runs locally on each mobile device based on available meta-data information, which consists of the user-item rating matrix and contact history.

We denote the element $m_{u,i}$ as the rating of item $i$ by user $u$ at any given time. The status of $m_{u,i}$ can be either **rated**, **predicted** or **unpredictable**. A rating $m_{u,i}$ is rated if $i$ has be transferred to and rated by $u$, and the rating can be either 1 or 0. A rating $m_{u,i}$ is predicted if it has not been rated yet, but the rank $R_{u,i}$ (see Equation 2) can be computed. The predicted rating is 1 if $i$ is among the top-$k$ item according to $R_{u,i}$ for user $u$, and 0 otherwise. A rating $m_{u,i}$ is said to be *correct* if the predicted rating matches the user rating eventually.

Recall that there are two naive methods to pick an item to transfer to another device. The first method, considering only item recall, picks a predicted item that gives the highest rank $R_{u,i}$ to maximize the probability that the rating $m_{u,i}$ is correct and positive. The second method considers only the prediction coverage, and picks a predicted item such that *if* the item is rated, then the number of unpredictable items becoming predictable is maximal.

Table 3. List of variables used.

| Notation | Description |
|----------|-------------|
| $n$ | number of users |
| $g_i^+$ | number of positive prediction for item $i$ currently |
| $r_i^+$ | number of correctly predicted positive ratings for item $i$ currently |
| $\Omega_i$ | random variable for number of correct positive ratings for item $i$ when all users have rated $i$ |
| $\sigma_q(i)$ | the queue position of item $i$ at node $q$ |
| $B$ | average device contact capacity |
| $\lambda$ | average device contact rate |
| $H_i$ | set of devices with item $i$ |

To consider both recall and coverage, we consider the following metric: for an item $i$, we are interested in the number of correct positive prediction for $i$ eventually, i.e., when $i$ has been rated by all users. Before $i$ is rated by all users, this quantity is considered as a random variable, denoted as $\Omega_i$. At any round $t$, we know the current number of correct positive rating for $i$, denoted $r_i^+$. We also know is the number of positive predictions for item $i$, $g_i^+$. Ideally, we would like the following inequality to be true:

$$\Omega_i > r_i^+ + g_i^+,$$

i.e., all the positive predictions for $i$ are correct, and there are additional new postive ratings for $i$. The key question is thus to estimate the probability that the above condition is true if $i$ is transferred.

In the following, we present approximations on the potential positive ratings for an item and the probability of delivery of items with positive ratings to the users. The goal is to derive approximations that can be used as input to guide and motivate the design of *CoFiGel*.

A list of variables used in the rest of this section is given in Table 3.

### 4.2 Potential Gain in Positive Ratings

First, we derive an equation to bound $P\{\Omega_i > g_i^+ + r_i^+\}$, the probability that the number of correct positive predictions for item $i$ would increase if $i$ is transferred.

**Theorem 1.**

$$Pr\{\Omega_i > r_i^+ + g_i^+\} \leq \min\left\{1, e^{\frac{r_i^+ E[\Omega_i]}{n - r_i^+}}\left(1 - \frac{r_i^+}{n}\right)^{r_i^+ + g_i^+}\right\} \qquad (3)$$

*Proof.* Let $\Omega_{u,i}$ be the random variable that takes a value of 1 if the predicted rating for the user $u$ for item $i$ is correct and positive, and a value of 0 otherwise. Let $p_{u,i}$ be the probability that the prediction for the user $u$ on item $i$ is correct and positive, i.e., $p_{u,i} = Pr\{\Omega_{u,i} = 1\}$.

As $\Omega_i = \sum_{u \in U} \Omega_{u,i}$, we can write $Pr\{\Omega_i > x\}$ as

$$Pr\{\Omega_i > x\} \le e^{-c \cdot x} \prod_{u \in U} \left(p_{u,i} e^c + 1 - p_{u,i}\right)$$

for any $c > 0$ and $x > 0$ (Lemma 1 of [6]).

Let $c = \ln\left(\frac{n}{n - r_i^+}\right)$ and $x = r_i^+ + g_i^+$, assuming that $0 < r_i^+ < n$. We rewrite the equation as follow:

$$Pr\{\Omega_i > r_i^+ + g_i^+\} \le$$
$$\left(1 - \frac{r_i^+}{n}\right)^{r_i^+ + g_i^+} \prod_{u \in U} \left(\frac{n p_{u,i}}{n - r_i^+} + 1 - p_{u,i}\right) \tag{4}$$

From Taylor series expansion, we know that

$$\frac{n p_{u,i}}{n - r_i^+} + 1 - p_{u,i} = 1 + \frac{r_i^+ p_{u,i}}{n - r_i^+} \le e^{\frac{r_i^+ p_{u,i}}{n - r_i^+}}$$

Since

$$\prod_{u \in U} e^{\frac{r_i^+ p_{u,i}}{n - r_i^+}} = e^{\frac{r_i^+ \sum_{u \in U} p_{u,i}}{n - r_i^+}} = e^{\frac{r_i^+ E[\Omega^i]}{n - r_i^+}}$$

We have

$$Pr\{\Omega_i > r_i^+ + g_i^+\} \le e^{\frac{r_i^+ E[\Omega^i]}{n - r_i^+}} \left(1 - \frac{r_i^+}{n}\right)^{r_i^+ + g_i^+}$$

We bound the probability to 1.

Some points to notes:

– $E[\Omega_i]$ is unknown and needs to be estimated. In the algorithm implementation, we approximate $E[\Omega_i]$ using $r_i^+$. Once we substitute $E[\Omega_i]$ with $r_i^+$, the right hand side of Equation 3 can be computed using the parameters $r_i^+$ and $g_i^+$ which are available locally. Further, to bootstrap the system, we set initialize $r_i^+$ to a small value (we set it to 1% of all total users in our simulation).
– The concentration bound decreases as $g_i^+$ increases: as we made more positive predictions for item $i$, the probability that the predictions are all correct decreases.
– The concentration bound increases for large $r_i^+$ as $r_i^+$ increases. If an item has lots of correct positive ratings, then the chances that current positive predictions are correct increases.

### 4.3 Estimation of Delivery Probability

In the previous section, we only consider the effect of transfer an item $i$ on the CF algorithm, but not how long it takes. For the ratings and items to be useful, the item should reach a user before some time deadline. We estimate the probability it takes to delivery an item $i$ late after the deadline in this section.

Consider a DTN with exponentially distributed inter-contact time and infinite node buffer capacity. Let $\lambda$ be the average contact rate (number of contacts in unit time) and $B$ be the average contact capacity (duration of contact in time units) of the system. Let $H_i$ be the set of users having item $i$ and $N_i$ be the set of users that $i$ should be delivered to.

Let $Y_i$ be the random variable representing the time when $i$ is delivered to *all* users in $N_i$ and $Z_i$ be the time when $i$ is delivered to *any* user in $N_i$. Obviously,

$$Y_i \leq \sum_{u \in N_i} Z_i \tag{5}$$

By Markov inequality and Equation 5, we can find the probability that a given item $i$ is delivered after a given time $t$:

$$Pr\{Y_i \geq t\} \leq \frac{E[Y_i]}{t} \leq \frac{E[\sum_{u \in N_i} Z_i]}{t} = \frac{|N_i|E[Z_i]}{t} \tag{6}$$

The next step is to approximate $E[Z_i]$. An item $i$ is delivered to a user $u$ only if $u$ comes into contact with a user in $H_i$ for a long enough time for $i$ to be transferred. Furthermore, we consider $i$ as a useful delivery only when this occurs the first time. If we let $M_{i,u}$ be the random variable representing the time when user $u$ meets with a user in $N_i$ and transfers $i$ to that user, then $Z_i = \min_{u \in H_i}\{M_{i,u}\}$.

Let $\mu_{i,u}$ be $E[M_{i,u}]$, and $\mu_i$ be the mean of $\mu_{i,u}$ over $u \in H_i$. i.e.,

$$\mu_i = \frac{1}{|H_i|} \sum_{u \in H_i} \mu_{i,u}$$

From Bertsimas et al. [7], we know that the expected value of the first order statistic, $Z_i$, is bounded by the average of the individual means of the random variables of the order statistic:

$$E[Z_i] \leq \mu_i$$

and, therefore, from Equation 6:

$$Pr\{Y_i \geq t\} \leq \frac{|N_i|\mu_i}{t} \tag{7}$$

We now explain how we can estimate $\mu_i$ and bound $Pr\{Y_i \geq t\}$. Upon device contact, each device exchanges a matrix $\sigma$, where each element $\sigma_{i,v}$ is the last known position (in bytes) of item $i$ in user $v$'s transfer queue for $v \in H_i$. We approximate the system behavior by assuming that scheduling is FIFO. The time user $v$ meet any user $u$ before transferring item $i$ to $u$ is estimated with $\frac{\sigma_{i,v}}{B\lambda}$. $\mu_i$ is then approximated as

$$\mu_i = \frac{1}{B\lambda|H_i|} \sum_{v \in H_i} \sigma_{i,v} \tag{8}$$

Putting the results together, from Equations 8 and 7, and bounding the probability to 1,

$$Pr\{Y_i \geq t\} > 1 - \min\{1, \frac{|N_i|}{B\lambda t|H_i|} \sum_{v \in H_i} \sigma_{i,v}\} \tag{9}$$

The values $|N_i|$ and $|H_i|$ can be obtained from the user-item rating matrix being exchanged, and $B$ and $\lambda$ can be estimated locally by keeping track of contact history.

We note the following in Equation 9. The likelihood of not meeting the deadline for an item $i$ increases if (i) $|N_i|/|H_i|$, the ratio of number of nodes not having $i$ to having $i$ increases, (ii) $B\lambda t$, the contact opportunity until time $t$, decreases, and (iii) $\sigma_{i,v}$, the waiting time for $i$ in the transfer queue increases.

### 4.4 Algorithm

Now we can present the workings of CoFiGel based on the results derived from previous sections. At each device, CoFiGel decides which item to transmit by computing a utility $U_i$, which incoperates the number of positive ratings (rated or predicted) for $i$, the probability of gain in ratings, and the probability of delivery within the deadline:

$$U_i = (g_i^+ + r_i^+) \cdot G_i \cdot D_i \tag{10}$$

where $G_i$ is the right-hand-side of Equation 3, and $D_i$ is the right-hand-side of Equation 9.

The utility increases if either (i) the total number of correctly predicted positive ratings we get eventually $(g_i^+ + r_i^+)$, increases (ii) the likelihood of the number of correct predictions increases $(G_i)$, or (iii) the likelihood of delivering an item within the deadline $t$ increases.

Note that since the bounds provided are very loose, we do not expect these computed utilities to reflect the true value of the rating gain. For the purpose of the scheduling, however, only the relatively ordering is important. Items with larger utilities are transferred first. We will shown in the evaluation that the heuristics used is indeed sufficient to provide good results.

## 5 Simulation Evaluation

### 5.1 Simulation Setting

In order to evaluate *CoFiGel*, we use the MovieLens data set (http://movielens.umn.edu) as the underlying user ratings. The data set chosen has 100K ratings, 943 users and 1682 items.

For mobility traces we have chosen to use the RollerNet trace ([11]) and the San Francisco taxi trace ([4]) or SanCab trace.

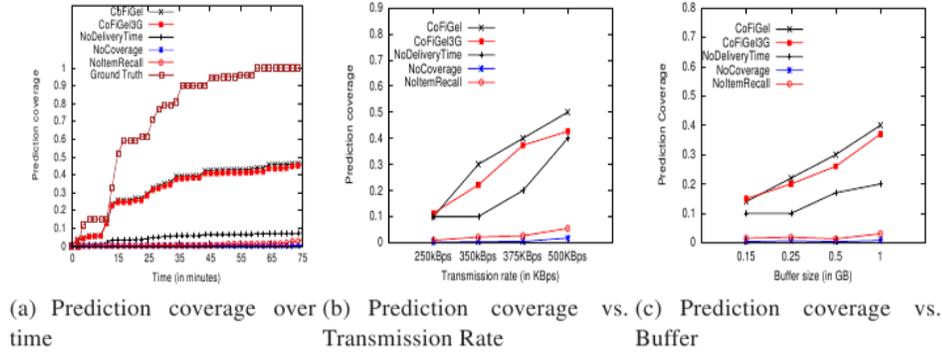

(a) Prediction coverage over time
(b) Prediction coverage vs. Transmission Rate
(c) Prediction coverage vs. Buffer

**Fig. 1.** RollerNet trace (total ratings = 11536)

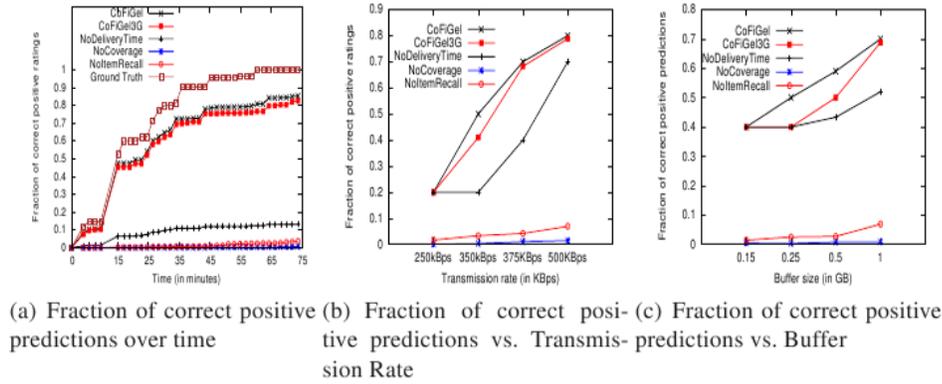

(a) Fraction of correct positive predictions over time
(b) Fraction of correct positive predictions vs. Transmission Rate
(c) Fraction of correct positive predictions vs. Buffer

**Fig. 2.** RollerNet trace (total positive ratings = 6400)

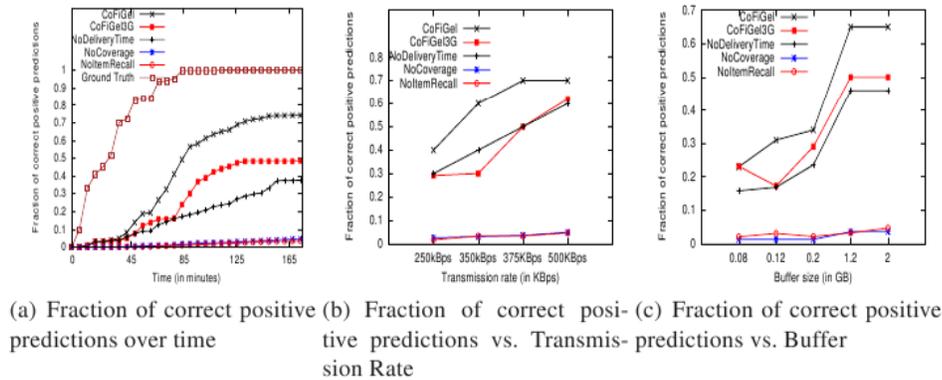

(a) Fraction of correct positive predictions over time
(b) Fraction of correct positive predictions vs. Transmission Rate
(c) Fraction of correct positive predictions vs. Buffer

**Fig. 3.** SanCab trace (total positive ratings = 5400)

**Table 4.** Simulation Parameters

| Parameter | SanCab | RollerNet |
|---|---|---|
| Number of Publisher Nodes | 22 | 10 |
| Number of Subscriber Nodes | 56 | 30 |
| (Item publisher rate)/publisher | 20 items/Hr | 40 items/Hr |
| Simulation duration | 6 Hrs | Approx.3 Hrs |
| Item size | 11MB | 15MB |
| Default buffer size | 2GB | 1GB |
| Default contact bandwidth | 3Mbps (375KBps) | 3Mbps |
| Item Lifetime | 2 hours | 1 hour 15 min |
| Warmup time | 1 Hr | 1 Hr |
| Cool down time | 1 Hr | 0.5 Hr |

The RollerNet trace is an example of human mobility and the data set consists of about 60 Bluetooth devices carried by groups of roller bladers in a roller tour over three hours. The average contact duration is 22 seconds and the average number of contact per node over 3 hours is 501.

The San Francisco taxi (SanCab) trace is representative of a vehicular DTN environment. The data set consists of GPS coordinates of about 500 taxis in the San Francisco bay area over a one month period. We selected a 6 hour interval for our simulation. By assuming a fixed communication range of 50m, we derive a communication pattern that has average contact duration of 73 seconds and the average number of contact per node over 6 hours is 213.

Video size of user generated content such as those found in popular sites like YouTube is 25MB or less (98% of videos are 25MB or less [9]). We choose data size of 11MB for the taxi trace and 15MB for RollerNet. The relatively larger item size for RollerNet is due to the large number of contacts per node. The buffer size and item generation rate are similarly adjusted to ensure sufficient loading in the system.

As some nodes in the trace have very limited contacts with the rest of the trace, we avoid selecting these nodes as the publisher or subscriber nodes (though they can still act as relay nodes). These nodes are identified as nodes which do not have sufficient number of node contacts and contact bandwidth to support meaningful data exchange. After removing these nodes, 22 publisher and 56 subscribers were chosen for the Taxi trace and 10 publishers and 30 subscribers were chosen for the RollerNet trace.

In order to reduce simulation time, we reduce the MovieLens data set selected by randomly choosing 900 items(movies) and 500 users from the original data set. All user-item ratings associated with these chosen user-item pairs from the original dataset are also included.

Finally, as the rating data set and the mobility trace are generated independently, we map the rating data to the mobility trace in the following way:

1. Every item in the reduced data set is randomly assigned to a publisher node in the mobility trace. This node will act as the publisher for the item.

2. Every user in the reduced data set is randomly mapped onto a mobile subscriber node. The actual user-item rating is known only when the item reaches the given mobile node where the user is located.

The settings in Table 4 are used as default unless otherwise specified. Each simulation point is run at least 3 times with different random seeds.

The performance objectives used are prediction coverage, precision, recall and number of satisfied users and latency, as described in Section 3.4. We compare the performance of *CoFiGel* with four other algorithms, namely:

1. A scheme that knows the ground-truth of data available. The ground-truth is available from the MovieLens data set. This scheme provides the actual rating coverage and gives an upper bound on the system performance. This scheme is used only in the coverage comparison since ground-truth is not applicable in the user satisfaction evaluation.

2. An epidemic-based algorithm that is similar to *CoFiGel* except that it does not take into account contact history and time constraints. We called this algorithm **NoDeliveryTime**. The performance difference between **NoDeliveryTime** and *CoFiGel* indicates the improvement provided by exploiting contact history.

3. An algorithm that uses only the rating information available. This is referred to as **NoCoverage**. The ratings of the items are predicted using the MCF, but the rating update and the potential coverage increase is not considered. By using only limited rating information, **NoCoverage** is expected to perform the worst.

4. An algorithm which tries to schedule an item so as to acquire prediction coverage of hitherto unrated users and to satisfy as many more users as possible. This is called the **NoItemRecall**. While this approach also uses contact history, it does not perform multi-round predictions as in the case of *CoFiGel*. It only acts using the current rating information.

5. **CoFiGel3G** is a modification of *CoFiGel* such that it uses the cellular network to upload/download ratings and a central server to run the MCF. However, the data are still sent over the DTN. By exploiting the cellular network as control channel, ratings information propagate quickly among the nodes and is always up-to-date. However, it is important to note that faster rating propagation does not always translate to higher rating coverage. This is because an actual rating can only be discovered after an user has access to the actual video and provides the rating.

## 5.2 Coverage

In this section, we evaluate the performance of *CoFiGel* and the other algorithms in terms of prediction coverage, a commonly used metric for MCF. In addition, we also measured the fraction of correctly predicted positive (or FCPP) items which looks at the ratio of correctly predicted positive item to the total of number of positive ratings rather over all ratings. Given that we are simulating a DTN environment, we felt that FCPP provides a better gauge for what is achievable by good algorithms in more challenging environments. Due to space constraints, we will only show the results for prediction

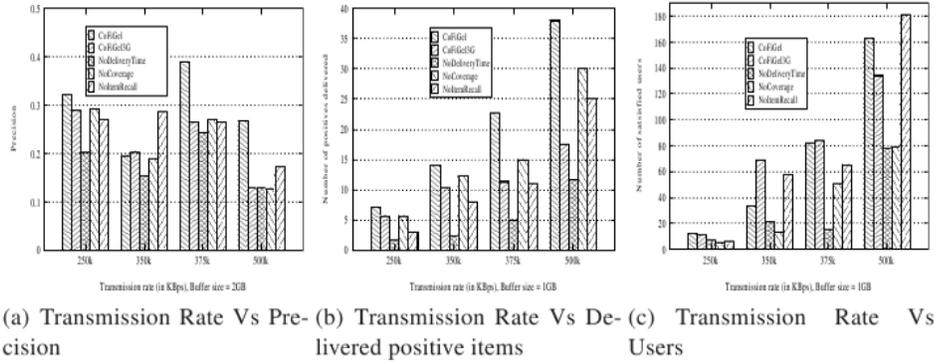

(a) Transmission Rate Vs Pre-   (b) Transmission Rate Vs De-   (c) Transmission Rate Vs
cision                          livered positive items          Users coverage for RollerNet (Figure 1(c)), but FCPP for both RollerNet (Figure 2(c)) and SanCab (Figure 3(c)).

**Fig. 4.** RollerNet trace

Figures 1(a), 2(a) and 3(a) show how positive ratings increase over time. The actual number of ratings for the items published so far (*Ground-Truth*) are shown to illustrate the best possible outcomes over time. The results show that *CoFiGel* and *CoFiGel3G* have the best performance. In the SanCab trace, *CoFiGel* performs even better than *CoFiGel3G*. In terms of overall ratings, CoFiGel discovers 45% of the ratings in the RollerNet trace. In terms of FCPP, *CoFiGel* discovers 84% and 74% of the positive ratings in the RollerNet and SanCab traces respectively. The result is better in the Roller-Net case due to higher contact rate and capacity. In fact, with RollerNet, the performance of *CoFiGel* measured using FCPP closely matches the actual ratings in the first 15 minutes and the gap remains small throughout the simulation.

In the results shown, it can be clearly observed that *CoFiGel* has the best performance, followed by *CoFiGel3G*. This result can be somewhat surprising since *CoFiGel3G* uses the same algorithm as *CoFiGel* but uses the control (3G) channel for centralized rating computation and sharing. We explain the result as follows. Since *CoFiGel3G* performs centralized rating, the rating matrix gets updated much faster. This fast rating update has the (unintended) consequence that the variable $G_i$ in Equation 10 approaches the value of 1.0 much faster than the case for *CoFiGel*. As the value of $G_i$ gets close to 1 and saturates around this value, this variable becomes useless in term of providing information for relative ranking to decide which video data item is more important. However, since propagation of video data item lags behind rating data, the loss of this rating information results in *CoFiGel3G* performing worse than *CoFiGel*.

For the RollerNet trace simulation with default settings, the higher contact rate and capacity turn out to have adverse effect on **NoDeliveryTime**, **NoCoverage** and **NoItemRecall**, since each algorithm only looks at one aspect of the problem. In terms of FCPP, **NoDeliveryTime** discovers 13% of the positive ratings, while **NoCoverage** discovers 0.6% or less of the positive ratings and **NoItemRecall** discovers around 1%.

For the SanCab trace, **NoDeliveryTime**'s performance is similar to *CoFiGel* in the early part of the experiment but since it does not take deadline into account, its performance degrades with time. As expected, as **NoCoverage** uses only basic rating, it performs very badly. This is also the case with **NoItemRecall**. Overall, in terms of FCPP, for the SanCab trace, **NoDeliveryTime** discovers 38% of the ratings, while **No-Coverage** and **NoItemRecall** discover 4% or less.

For both traces, the coverage for the **NoCoverage** is very low, showing that it is important to take into account additional information beyond ratings. Figures 1(b), 2(b) and 3(b) show how coverage varies with transmission rate. While increase in contact capacity results in increased coverage because more items get rated, *CoFiGel* is able to exploit the increase in transmission rate much better than **NoDeliveryTime**, **NoCover-age** and **NoItemRecall**. For the SanCab trace, in terms of FCPP, relative to *CoFiGel*, **NoDeliveryTime** discovers 17% to 46% less ratings while **NoCoverage** and **NoItem-Recall** discover less than 5% of the ratings consistently.

Similar behavior is observed in the RollerNet trace. In the results shown, *CoFiGel* performs better than **NoDeliveryTime** by up to 105% and discovers at least 50 times more ratings than **NoCoverage** and **NoItemRecall** consistently. In general, more im-provement comes from taking into account rating coverage gain (from **NoCoverage** to **NoDeliveryTime**) than taking into account contact history. The effort by **NoItem-Recall** to increase the number of user ratings is also ineffective due to the absence of rating gain which is capitalized by *CoFiGel*. Nevertheless, substantial improvement is still observed between **NoDeliveryTime** and *CoFiGel*.

The performance with respect to different buffer sizes is shown in Figures 1(c), 2(c) and 3(c). For the SanCab trace, buffer size is varied from $80MB$ to $2GB$. The performance of *CoFiGel* is better than **NoDeliveryTime** by 41% to 83%. Improvement of *CoFiGel* over **NoCoverage** is about 17 times. For the RollerNet trace, there are two observations. First, for very small buffer size of less than 150MB, very few items make it to the next hop and hence, the FCPP remains same for *CoFiGel* and **NoDeliveryTime**. FCPP of *CoFiGel* is higher than **NoDeliveryTime** by up to 36% and for **NoCoverage** by 50 to 60 times.

### 5.3 User Satisfaction

While coverage indicates the predictive power of the system, the actual user satisfac-tion has to be measured by looking at how many items reach users that like them. In order to ensure that the nodes have accumulated enough training data before making the measurement, for the SanCab trace, we consider items generated after the first and before the fifth hour. The first hour serves as the training phase, while the last hour is ignored to make sure that items generated later in the trace do not bias the measurement. Similarly, the training phase for RollerNet is 1.5 hours and items generated during last half hour of trace are ignored.

Figure 4(a) shows the results for precision of items reaching the users. It is clear that *CoFiGel* performs very well, except for one case (350K), it has the highest precision. In addition, note that even though **NoItemRecall** has a higher precision, from the results in the previous section, it has very low coverage. Due to the disconnected nature of DTN and the large number of data items and users available, it is also useful to look user utility in two other ways.

First, we look at the average number of positively rated items that reach any user. The result is shown in Figures 4(b) and 5(a). *CoFiGel* clearly outperforms the other two algorithms by a very large margin once the bandwidth exceeds some threshold required for data dissemination. For the SanCab trace, at the highest transmission rate experimented, *CoFiGel* delivers up to 59% more useful items than **NoDeliveryTime** and 245% more useful items than **NoCoverage**. In the case of the RollerNet trace, the improvements are 117% and 225% respectively.

Another way we measure recall is to look at the number of users who have received at least one useful item. The result is shown in Figures 4(c) and 5(b). Again, *CoFiGel* performs well, in particular, at higher bandwidth. At 4Mbps, *CoFiGel* delivers twice as many useful items to users than **NoCoverage** and **NoDeliveryTime** for both traces. At 2Mbps, both *CoFiGel3G* and **NoItemRecall** outperform *CoFiGel* in the SanCab trace. The much more sparse contact interval and lower contact bandwidth may have help to slow down rating matrix propagation enough such that *CoFiGel3G* is able to fully exploit the benefit of fasting rating propagation without value saturation.

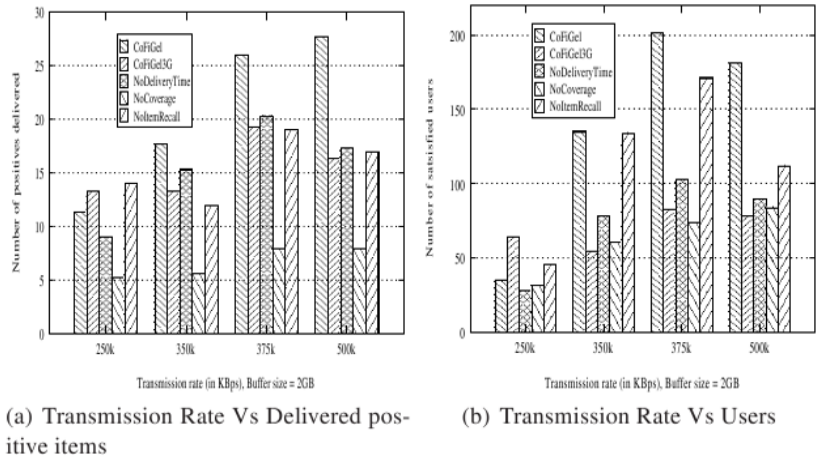

(a) Transmission Rate Vs Delivered positive items

(b) Transmission Rate Vs Users

**Fig. 5.** SanCab trace

## 6 Conclusion

We have presented *CoFiGel*, a novel approach that combines collaborative filtering and DTN routing so as to enable content distribution in a distributed environment with intermittent connectivity. It is designed for sharing of locally stored contents that have *spatial and temporal relationships*. *CoFiGel* has two key components. First, it estimates the potential gain in prediction coverage if an item is scheduled. Second, it estimates the time needed to deliver the ratings. Our analysis allows us to derive approximations that are used as input to the utility computation in *CoFiGel* based on only locally estimated parameters.

Simulation results show that *CoFiGel* performs well. It is able to discover substantially more ratings and is able to deliver much more items that are rated positively by the users than baseline algorithms that do not take into account rating gain or mobility pattern.

## References

1. O2 Germany admits network meltdown; smartphones blamed, http://www.fiercewireless.com/europe/story/o2-germany-admits-network-meltdown-smartphones-blamed/2011-11-23
2. 3G to WiFi offload in the pipes for Singapore, http://www.telecomasia.net/content/3g-wifi-offload-pipes-singapore
3. MovieLens Dataset, http://www.grouplens.org/node/73
4. Michal Piorkowski., Natasa Sarafijanovic-Djukic., Matthias Grossglauser.: CRAWDAD data set epfl/mobility (v. 2009-02-24). http://www.grouplens.org/node/73
5. Huang, Junxian., Xu, Qiang., Tiwana, Birjodh., Mao, Z. Morley., Zhang, Ming., Bahl, Paramvir.: Anatomizing Application Performance Differences on Smartphones. In: MobiSys(2010), pp. 165–178
6. Diaz, J., Petit, J., Serna, M.: A Guide to Concentration Bounds. In: Handbook on Randomized Computing II, Combinatorial Optimization, vol. 9, pp. 457–507. Springer (2001)
7. Bertsimas, Dimitris., Natarajan, Karthik., Teo, Chung-Piaw.: Tight Bounds On Expected Order Statistics. In: Probabability in the Engineering and INformational Sciences, vol. 20, pp. 667–686. Cambridge University Press (2006)
8. Su, Xiaoyuan., Khoshgoftaar, Taghi M.: A survey of collaborative filtering techniques. In: Advances in Artificial Intelligence, vol. 2009, Hindawi Publishing Corp. (2009)
9. Xu, Cheng., Cameron, Dale., Jiangchuan, Liu.: Statistics and Social Network of YouTube Videos. In: IWQoS(2008), pp. 229–238
10. Vidal, José M.: A Protocol for a Distributed Recommender System. In: Trusting Agents for Trusting Electronic Societies. LNCS, vol. 3577, pp. 200–217. Springer,Heidelberg (2005)
11. Tournoux, P. U., Leguay, J., Benbadis, F., Conan, V., Dias de Amorim, M., Whitbeck, J.: The Accordion Phenomenon: Analysis, Characterization, and Impact on DTN Routing. In: INFOCOM(2009), pp. 1116–1124
12. Berkovsky, Shlomo., Kuflik, Tsvi., Ricci, Francesco.: Distributed collaborative filtering with domain specialization. In: RecSys(2007), pp. 33–40
13. Wang, Jun., Pouwelse, Johan., Lagendijk, Reginald L., Reinders, Marcel J. T.: Distributed collaborative filtering for peer-to-peer file sharing systems. In: SAC(2006), pp. 1026–1030
14. Hu, Liang., Le Boudec, Jean-Yves., Vojnovic, Milan.: Optimal Channel Choice for Collaborative Ad-Hoc Dissemination. In: INFOCOM(2010), pp. 614–622
15. Balasubramanian, N., Balasubramanian, A., Venkataramani, A.: Energy Consumption in Mobile Phones: A Measurement Study and Implications for Network Applications. In: IMC(2009), pp. 280–293
16. Balasubramanian, A., Levine, B., Venkataramani, A.: DTN routing as a resource allocation problem. In: SIGCOMM(2007), pp. 373–384
17. Ioannidis, Stratis., Chaintreau, Augustin., Massoulie, Laurent.: Optimal and Scalable Distribution of Content Updates over a Mobile Social Network. In: INFOCOM(2009), pp. 1422–1430
18. Altman, E., Nain, P., Bermond, J.: Distributed Storage Management of Evolving Files in Delay Tolerant Ad Hoc Networks. In: INFOCOM(2009), pp. 1431–1439
19. K. C.-J. Lin., C.-W. Chen., C.-F. Chou.: Preference-Aware Content Dissemination in Opportunistic Mobile Social Networks. In: INFOCOM(2012), pp. 1960–1968

20. Karypis, George.: Evaluation of Item-Based Top-N Recommendation Algorithms. In: CIKM(2001), pp. 247–254
21. Reich, Joshua., Chaintreau, Augustin.: The Age of Impatience: Optimal Replication Schemes for Opportunistic Networks. In: CoNEXT(2009), pp. 85–96
22. Quercia, Daniele., Hailes, Stephen., Capra, Licia.: MobiRate: making mobile raters stick to their word. In: UbiComp(2008), pp. 212–221
23. Lo Giusto, Giacomo., Mashhadi, Afra J., Capra, Licia.: Folksonomy-based reasoning for content dissemination in mobile settings. In: CHANTS(2010), pp. 39–46
24. Lemire, Daniel., Maclachlan, Anna.: Slope One Predictors for Online Rating-Based Collaborative Filtering. In: SDM(2005)
25. Nordström, Erik., Gunningberg, Per., Rohner, Christian.: A Search-based Network Architecture for Mobile Devices. Technical Rep.2009-003, Uppasala University, Sweden (2009)
26. Krifa, A., Baraka, C., Spyropoulos, T.: Optimal Buffer Management Policies for Delay Tolerant Networks. In: SECON(2008), pp. 260–268
27. Linden, G., Smith, B., York, J.: Amazon.com recommendations: item-to-item collaborative filtering. In: Internet Computing, vol. 7, pp. 76–80. IEEE (2003)
28. Miller, Bradley N., Konstan, Joseph A., Riedl, John.: PocketLens: Toward a Personal Recommender System. In: Transactions on Information Systems, vol. 22, pp. 437–476. ACM (2004)
29. Li, Yong., Qian, Mengjiong., Jin, Depeng., Su, Li., Zeng, Lieguang.: Adaptive optimal buffer management policies for realistic DTN. In: GLOBECOM(2009), pp. 1–5
30. Lee, Kyunghan., Yi, Yung., Jeong, Jaeseong., Won, Hyungsuk., Rhee, Injong., Chong, Song.: Max-Contribution: On Optimal Resource Allocation in Delay Tolerant Networks. In: INFOCOM(2010), pp. 1136–1144